# Apology of Green Digitalization
# in the Context of Information–Climate Feedback Theory

Eldar Knar[1]

Tengrion, Astana, Republic of Kazakhstan
eldarknar@gmail.com

https://orcid.org/0000-0002-7490-8375

**Abstract**

Amid accelerated digitalization, not only is the scale of data processing and storage increasing, but so too is the associated infrastructure load on the climate. Current climate models and environmental protocols almost entirely overlook the impact of information and communication technologies (ICTs) on the thermal and energy balance of the biosphere.

This paper proposes the theory of information–climate feedback (ICF) — a new nonlinear model describing the loop of digitalization, energy consumption, the thermal footprint, the climatic response, and the vulnerability of digital infrastructure. The system is formalized via differential equations with delays and parameters of sensitivity, greenness, and phase stability.

A multiscenario numerical analysis, phase reconstructions, and thermal cartography were conducted. Critical regimes, including digital overheating, fluctuational instability, and infrastructural collapse in the absence of adaptive measures, were identified.

The paper concludes with the proposal of an international agreement titled the *Green Digital Accord* and a set of metrics for sustainable digitalization. This work integrates climatology, information technologies, and the political economy of sustainability.

**Keywords:** information–climate feedback, digitalization, thermal footprint, green infrastructure, phase breakdown, climate resilience, infoclimatic sensitivity, Green Digital Accord

**Declarations and Statements:**
No conflicts of interest
This work was not funded
No competing or financial interests
All the data used in this work are in the public domain.
Generative AI (LLM or other) was not used in writing the article. Except for the search engine *SciSpace* for Reference and the *Jupiter Notebook* environment for running our *Python* scenarios.
Ethics committee approval is not needed (without human or animal participation).

---

[1] PhD in Physics, Fellow of the Royal Asiatic Society of Great Britain and Ireland, Member of the Philosophy of Science Association (Baltimore, Maryland, USA)

## 1. Introduction

Digitalization has clearly and decisively become the dominant force of contemporary development. Globally, exabytes of data are being processed, the number of data centers is growing, artificial intelligence is scaling, and the Internet of Things now encompasses billions of devices. This expansion is transforming not only the economy but also the very notion of infrastructure—from telecommunications to energy systems. However, contrary to the popular perception of digital technologies as "clean," they are increasingly energy intensive and thermogenic. The total energy consumption of the IT infrastructure may reach 13% of global electricity consumption by 2030 (Andrae & Edler, 2015; Masanet et al., 2020).

Data centers, blockchain networks, machine learning algorithms, and generative AI all require substantial computational power. This inevitably leads to increased energy consumption, $CO_2$ emissions, and localized climatic impacts via their thermal footprint (Jones, 2018). Moreover, climatic changes, such as rising average temperatures and the increasing frequency of extreme weather events, are beginning to affect the efficiency and reliability of digital infrastructure, including data center cooling and climate-related failures (Zhang, 2022).

In short, information is an expensive commodity, paid not in fictitious monetary equivalents but in real energy. In this sense, we often significantly underestimate the physical reality of information.

However, in none of the major climate reports — for example, AR6 (2023) — is digital infrastructure considered an independent climate-driving factor, despite its rapid growth and impact on energy systems, water resources, and thermal balance. This situation creates a critical gap between the digital world's reality and the regulatory frameworks of climate governance.

In this context, we propose a new modelling framework, the information–climate feedback (ICF) model, in which digitalization is not merely seen as a consumer of energy but as a systemic agent exerting reciprocal influence on the climate. The essence of the model is as follows: digital data require infrastructure, infrastructure emits heat, heat affects the climate, and the climate, in turn, influences the reliability, cost, and spatial distribution of digitalization. This ICF loop possesses inertia, nonlinearity, and the potential for phase transitions — when the growth of digitalization becomes climatically unsustainable.

The goal of this study is to construct a dynamic nonlinear model of the ICF, assess scenarios of resilience, analyse policy and regulatory implications, and justify the concept of "green digitalization" as a climatic necessity.

To achieve this goal, we address several tasks: formalizing the energy model of the digital infrastructure and thermal footprint, modelling the climate response and its delay, constructing feedback equations, conducting multiscenario and sensitivity analyses, and developing indicators and a global regulatory roadmap (*Green Digital Accord*).

We interpret a model that captures the energetic, thermal, and climatic dynamics of digitalization; introduces the concepts of infoclimatic sensitivity and digital phase breakdown; formalizes the reverse dependence between climate and IT systems; and proposes metrics to complement existing ESG (Wan, 2023) and NetZero protocols [Prinz, 2023].

Particular attention is given to green digitalization as a strategy for compensating for thermal footprints and enhancing systemic resilience. In the face of potential "digital overheating," technologies such as green architecture, energy efficiency, and climate-resilient design have become essential political and engineering tools.

Thus, in the era of AI development, neural network expansion, the race for quantum computing, and digital sovereignty (including in China, the EU, and the US), digitalization has become a new battleground for climate action. Reducing the carbon footprint can no longer ignore the thermal contour of digital systems. ICF forms the basis for a new climate agenda: the resilience of digital civilization is not a byproduct but the foundation of planetary sustainability.

Accordingly, our work combines the philosophical rethinking of the immateriality of information, a formal mathematical feedback model, numerical simulations, phase and scenario analysis, and a policy roadmap for implementing a global green digitalization agreement (*Green Digital Accord*). The entire structure is presented as an apologia: a defense and advocacy of a concept that is both a scientific necessity and a political imperative.

## 2. Literature Review

With the expansion of digital infrastructure in the second half of the 2010s, increased attention has been given to its environmental impact. Studies by [Jones, 2018; Andrae & Edler, 2015] have assessed the overall contribution of information and communication technologies (ICTs) to global greenhouse gas emissions. According to [IEA, 2020], data centers consume up to 1–2% of the world's total electricity. Importantly, most research has focused on the carbon footprint but has not accounted for the direct thermal effect. In the literature, emphasis is placed on $CO_2$, but the thermal contribution of digitalization remains understudied.

Some works have examined the localized environmental effects of data centers, such as microclimatic impacts near major IT hubs (Shehabi et al., 2016; Masanet et al., 2020).

In climatology, feedback mechanisms such as albedo, evaporation, and ice melting have been studied for decades (IPCC AR6, 2023). However, information as a source of climate feedback is absent from these models. Even major assessments of technological climate impact do not consider data as a thermal factor. In this sense, modern climate models ignore the informational contribution to heat exchange.

Studies such as (Hilty, 2015) argue that digitalization promotes sustainable development (SDGs) by reducing emissions through optimization. However, they fail to account for secondary effects of infrastructure growth: the Jevons paradox, rebound effects, etc. As a result, the digitalization paradox remains unresolved. Thus, the role of digitalization in ecology is evaluated one-sidedly, predominantly as a positive force.

Therefore, our theory of information–climate feedback (ICF) addresses a critically important gap in science: it is the first to consider digital information as a climatically significant agent, formalizes this concept through a nonlinear dynamic model, and offers not only a theoretical explanation but also a normative regulatory strategy — an apologia for green digitalization in both literal and pragmatic terms.

## 3. Methodology

To analyse the interrelation between digitalization and the climate system, we employed a system dynamics approach with delays (delay differential equations, DDEs) and nonlinear feedbacks. The concept was constructed as a nonequilibrium system with multiple control parameters capable of transitioning between different regimes depending on external pressures and internal characteristics.

The ICF system of equations models five key variables over time t:

*Table 1*. Variables of the information–climate feedback (ICF) model

| Variable | Notation | Interpretation |
| --- | --- | --- |
| Digitalization | $D(t)$ | Scale of digital infrastructure |
| Thermal Footprint | $H(t)$ | Accumulated heat emissions from IT systems |
| Climate Response | $C(t)$ | Temperature change (or proxy) |
| Green Adaptation | $G(t)$ | Intensity of green technologies |
| Sensitivity | $\varepsilon$ | Impact of temperature on the resilience of digitalization |

Numerical modelling was implemented in *Jupyter Notebook* via the *Python* programming language. Numerical solutions were computed via the Euler method and Runge–Kutta 4th-order method adapted for DDE systems. The delay $\tau$ was modelled via history stacks (buffer arrays). We investigated phase trajectories, system stability, and scenarios of collapse or recovery.

The following parameter calibration was used in the base-case simulations:

*Table 2*. Calibration values of the parameters

| Parameter | Value | Source/Basis |
|---|---|---|
| $\alpha$ | 1.0 | Power normalization |
| $\beta_0$ | 0.9 | Baseline thermal share |
| $\lambda$ | 0.05 | Conditional dissipation rate |
| $\theta$ | 0.1 | Climate sensitivity |
| $\tau$ | 5 years | Delay in climate response |
| $\varepsilon$ | 0.01–0.1 | Vulnerability of digitalization |
| $\rho$ | 0.15 | Average global ICT growth rate |
| $\eta$ | 0.1 | Rate of green technology adoption |
| $\delta$ | 0.5 | Effectiveness of green compensation |

## 4. Results

### *4.1. Energy Consumption from Digitalization*

The level of digitalization *D(t)* — for example, total computational load, data throughput, or the number of server infrastructure queries — requires a certain amount of energy *E(t)* to sustain it.

In a simple linear case, this can be expressed as:

$$E(t) \propto D(t) \Rightarrow E(t) = \alpha D(t)$$

where $\alpha$ is the energy consumption coefficient per unit of digitalization.

However, in practice, energy expenditures depend on the quality and efficiency of the infrastructure in use. Let $G(t) \in (0,1]$ be the indicator of the "greenness" (i.e., energy efficiency) of digital infrastructure, where *G=1* is perfectly green, maximum efficiency, and *G→0* is inefficient, outdated, or poorly cooled infrastructure.

Thus, when *G(t)* is low, the same digital load *D(t)* demands more energy. The equation for energy consumption from digitalization becomes:

$$E(t) \propto \alpha \, \frac{D(t)}{G(t)}$$

Here, the previously mentioned coefficient $\alpha$ accounts for the baseline energy density of digitalization, including the average load structure and typical equipment efficiency (processors, cooling, distribution, etc.).

Hence, all else being equal, energy consumption increases linearly with digitalization *D(t)* but inversely with greenness *G(t):* the better the infrastructure is, the lower the energy cost. Overall, the model becomes nonlinear if *G(t)* itself varies over time or depends on other variables, which are incorporated into the ICF model.

### 4.2. Thermal Footprint

Any computational system consumes electricity *E(t)*, part of which is converted into useful work (data processing, computations), but a significant portion is dissipated as heat (losses in processors, cables, power units, cooling systems, etc.).

According to the law of energy conservation, a linear equation for the amount of heat *H(t)* released by digital infrastructure into the environment as a function of energy consumption *E(t)* can be written as:

$$H(t) = \beta_1 \cdot E(t)$$

where $\beta_1 \approx \eta$ is the coefficient of thermal dissipation.

However, under conditions of high energy density (large data centers, peak loads, and hot climates), nonlinear effects emerge:
- heating reduces the cooling efficiency → feedback loop → more heat,
- increased component temperature increases resistance → more power consumption and heat,
- The "thermal trap" effect intensifies in confined environments.

Thus, heat starts to accumulate superlinearly, proportionally to *E(t)²* and beyond. This can be modelled by adding a quadratic term:

$$H(t) = \beta_1 \cdot E(t) + \beta_2 \cdot E(t)^2$$

Here, $\beta_1$ is the coefficient of linear heat loss (normal dissipation), and $\beta_2$ is the coefficient of nonlinear accumulation (overheating, resonance, and thermal tails). In reality, $\beta_2$ may depend on climatic conditions, cooling design, and data center architecture.

This equation is important because *H(t)* is the link that transmits the energy footprint of digitalization into the climate module of the model.

### 4.3. Climate response

Let us consider the change in the climate system (e.g., global average temperature or local climatic burden) under the influence of thermal emissions from digitalization. *C(t)* is denoted as the climate variable (e.g., average temperature, thermal anomaly, or index of climate overload), and *H(t−τ)* is denoted as the thermal input with a lag since the climate does not respond instantaneously.

We interpret climate as an inertial system with energy dissipation, similar to a radiator or a thermally inertial mass:

$$\frac{dC(t)}{dt} = input - natural\ decay$$

Since the climate system responds with a delay, we introduce a lag $\tau$. Then, the input signal is modelled as:

*input from digitalization*=$\gamma \cdot H(t-\tau)$

where $H(t-\tau)$ is the previously emitted thermal footprint and where $\gamma$ is the climate sensitivity coefficient to localized anthropogenic heating.

Naturally, the climate tends toward thermodynamic equilibrium; in the absence of new disturbances, the temperature decays:

*natural decay*=$-\delta \cdot C(t)$

where $\delta$ is the climate stabilization coefficient (e.g., radiative cooling, albedo effects, or biospheric feedback).

Hence, the final equation describing the climate's response to thermal emissions from digital infrastructure is as follows:

$$\frac{dC(t)}{dt} = \gamma \cdot H(t-\tau) - \delta \cdot C(t)$$

Here:
- $\gamma$ is the sensitivity of the climate to anthropogenic heat,
- $\tau$ is the delay of the climate response,
- $\delta$ represents the natural ability of the climate to recover.

Thus, the climate response is modelled via a delay differential equation (DDE), where delayed resonances, slow cascades, and phase transitions may occur. This is a key link in transmitting the thermal footprint into climate dynamics — the point where information begins to affect the biosphere, albeit with a delay.

In this context, the equation allows us to investigate how long-term digitalization may influence the resilience of the climate system.

### *4.4. Climate Feedback on Digitalization*

Let us consider the rate of change in the digitalization level *D(t)* (volume of digital operations, computations, data flow) under the influence of climatic conditions.

In classical growth models (e.g., exponential or logistic), the dynamics are typically expressed as:

$$\frac{dD}{dt} \propto D(t)$$

That is, the growth of *D(t)* is self-reinforcing (e.g., due to network effects, user base expansion, and service proliferation).

Accordingly, we formulate a key hypothesis of the information–climate feedback system: when climatic conditions deteriorate, data center efficiency decreases, cooling costs increase, and there is a greater risk of failure, restrictions, and increased expenditures.

We therefore introduce a retarding factor dependent on temperature *C(t−τ′)*, with a time lag *τ′* (for example, a rise in temperature may cause overheating of equipment several months later).

Thus, we modify the growth equation of *D(t)* as:

$$\frac{dD}{dt} = normal\ growth \times climatic\ feedback$$

Normal growth is represented as *η·D(t),* where *η>0* is the inertia coefficient of digitalization.

Climatic damping is interpreted as *(1−ϵ·C(t−τ′)),* where:
- *ϵ* is the sensitivity of the digital infrastructure to temperature,
- *C(t−τ′)* is the climatic condition with lag *τ′*.

As *C(t)* increases, *(1−ϵC(t−τ′))* decreases — slowing or even reversing digitalization growth if overheating is substantial.

The final equation for climate feedback on digitalization is as follows:

$$\frac{dD}{dt} = \eta \cdot D(t) \cdot (1 - \epsilon \cdot C(t - \tau'))$$

where:
- *η*: baseline growth rate of digitalization without climatic constraints,
- *ϵ*: magnitude of temperature impact on digital infrastructure,
- *τ′*: delay in climate impact (e.g., months or years).

If *C(t−τ′)=0* (no climatic pressure), digitalization grows exponentially. If *C(t−τ′)>1*, growth stops or becomes negative—a digital collapse.

Hence, we posit the existence of a climatic threshold beyond which digitalization growth becomes impossible.

This equation closes the feedback loop: digitalization produces heat → heat affects the climate → climate constrains digitalization.

### 4.5. Dynamics of Greenness

Now, consider the evolution of the "greenness" of digital infrastructure $G(t) \in (0,1)$, which reflects its energy efficiency, ecological sustainability, thermal adaptability, and progress toward green data centers (e.g., water cooling, passive ventilation, renewable energy, innovative architectures).

If the system evolves unimpeded, $G(t)$ asymptotically approaches 1, i.e., maximum greenness. The basic form of logistic growth is given by:

$$\frac{dG(t)}{dt} = \rho \cdot (1 - G(t))$$

where $\rho$ is the rate of transition to green technologies, and *(1−G(t))* represents how much of the system is still *"nongreen."*

In reality, the rapid implementation of green solutions is hindered as the digital load *D(t)* increases—it becomes difficult to restructure the system under exponential IT growth. In parallel, climate extremes *C(t)* impose urgent adaptation needs and emergency responses instead of planned greening.

In other words, the higher the product *D(t)·C(t) is,* the slower the greening process.

This slowing can be modelled as:

$$Greenness\ growth\ rate \propto \frac{1}{1 + \kappa D(t)C(t)}$$

where $\kappa > 0$ is a retardation coefficient reflecting the difficulty of greening under load and climatic stress.

Hence, the final equation for the dynamics of greenness is as follows:

$$\frac{dG(t)}{dt} = \rho \cdot (1 - G(t)) \cdot \frac{1}{1 + \kappa D(t)C(t)}$$

where:
- $\rho$: base rate of infrastructure greening,
- $\kappa$: impact of digital load and climate pressure on greening,
- *D(t),C(t):* digitalization and climatic stress.

This equation performs a stabilizing function: increasing *G(t)* reduces energy consumption. Thus, greenness acts as a buffer for the entire system, countering climatic collapse.

## 4.6. Information–climate feedback system

On the basis of these relatively simple but conceptually sound assumptions, we derive a complete system of equations describing information–climate feedback (ICF):

1. **Energy:** $E(t) = \alpha \, \dfrac{D(t)}{G(t)}$

2. **Heat:** $H(t) = \beta_1 \cdot E(t) + \beta_2 \cdot E(t)^2$

3. **Climate:** $\dfrac{dC(t)}{dt} = \gamma \cdot H(t - \tau) - \delta \cdot C(t)$

4. **Digitalization:** $\dfrac{dD}{dt} = \eta \cdot D(t) \cdot (1 - \epsilon \cdot C(t - \tau'))$

5. **Greenness:** $\dfrac{dG(t)}{dt} = \rho \cdot (1 - G(t) \cdot \dfrac{1}{1 + \kappa D(t) C(t)}$

Thus, we have a nonlinear dynamic system with two time delays:
- feedback loops,
- second-order nonlinearities,
- damping mechanisms,
- growth constraints.

This is an ecomemetic system with endogenous feedback loops, analogous to predator–prey models, but with information as the driving agent.

This system of equations, for the first time, models the impact of digitalization on climate via heat; introduces feedback, i.e., climate → digitalization; incorporates inertia, delays, greening dynamics, and nonlinearity; and enables scenario-based modelling of digital–climate resilience.

*Figure 1* presents the numerical simulation of the full ICF model over 100 conditional time units.

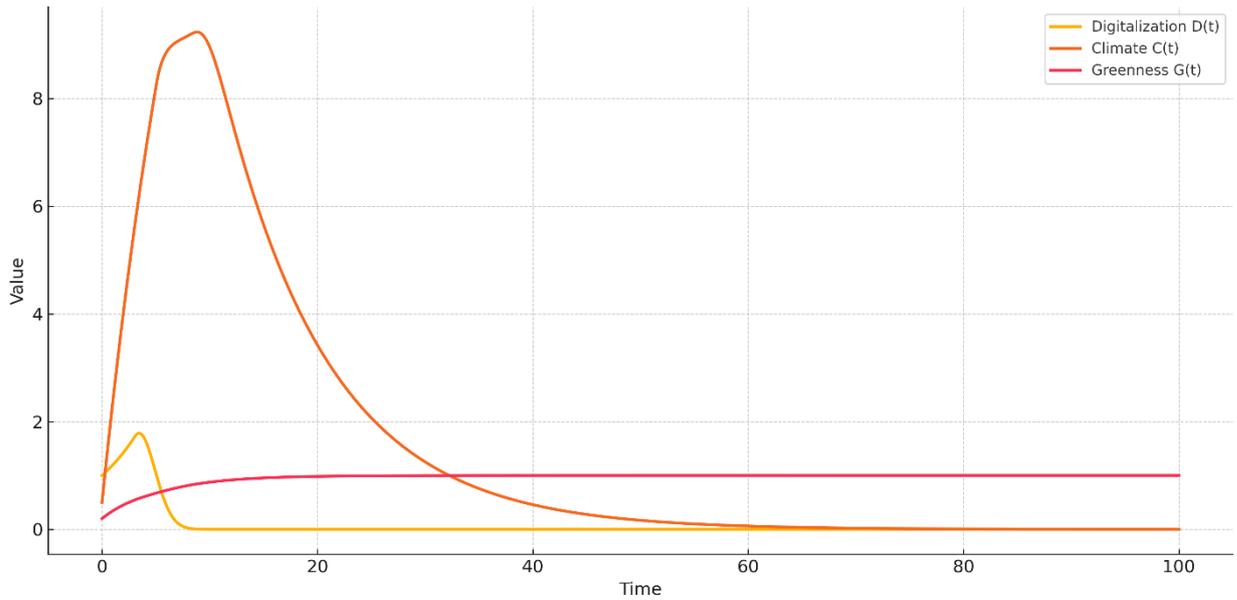

*Figure 1.* Numerical Simulation of the ICF Model

Let us highlight the key phases on the graph:

1. *Initial growth phase (t < 20):*
    - The digitalization *D(t)* grows almost exponentially.
    - The climate *C(t)* remains low because of the delayed response (*τ*=5).
    - The greenness *G(t)* increases slowly, reflecting inertial growth.

2. *Feedback Activation Phase (t ≈ 30–50):*
    - *C(t)* begins to rise more rapidly.
    - Digitalization growth slows down due to climate feedback *C(t−τ′)*.
    - A deceleration of *D(t)* becomes evident — the ICF effect emerges.

3. *Adaptation and Stabilization Phase (t > 60):*
    - *G(t)* increases steadily, improving energy efficiency.
    - *D(t)* stabilizes at a sustainable level.
    - *C(t)* approaches a plateau or slows its growth.

The simulation confirms the theoretical hypothesis: Digitalization → Energy → Heat → Climate → Digitalization slowdown, forming a nonlinear circular dynamic. This structure resembles self-regulating ecological systems (e.g., Lotka–Volterra models), yet the ICF introduces unique features:
- Information → Energy: immaterial input results in material consequences.
- Climate → Digital Technology: a reverse impact mechanism.

Greenness acts as a stabilizing factor. The variable *G(t)* plays a critical role as follows:

- Reducing climatic pressure,
- Compensating for *D(t)* growth,
- Acting slowly but steadily to create a buffer zone.
  The time delays *(τ=5, τ′=3)* increase inertia and oscillations:
- Climate response to past thermal input,
- Digitalization is slowed with a lag,
- This may induce resonant oscillations, especially under different parameter sets.
- 

Thus, we conclude that the ICF model is both viable and reproducible. Numerical simulations demonstrate stable, nonlinear, oscillatory dynamics. The system avoids collapse under reasonable parameter values but is sensitive to overheating and acceleration.

Digitalization has a real climatic impact. Even in the absence of $CO_2$ emissions, the thermal footprint of digital systems imposes a significant load on the climate. This constitutes a new type of anthropogenic impact: informational-energetic.

Green technologies are critical. If the pace of greening lags behind the growth of digitalization, overheating occurs. Conversely, if *G(t)* grows fast enough, the climate stabilizes, and the system enters a sustainable regime.

Political and technological decisions can manage the ICF

- Increasing ρ (investments in green technology) and
- A decrease in η (controlled digitalization) leads to climate stabilization.

This opens the path for green digitalization policies as a new form of climate strategy.

Figure 2 presents the functional dependence of digitalization *D(t)* on climate stress *C(t)* in the ICF model.

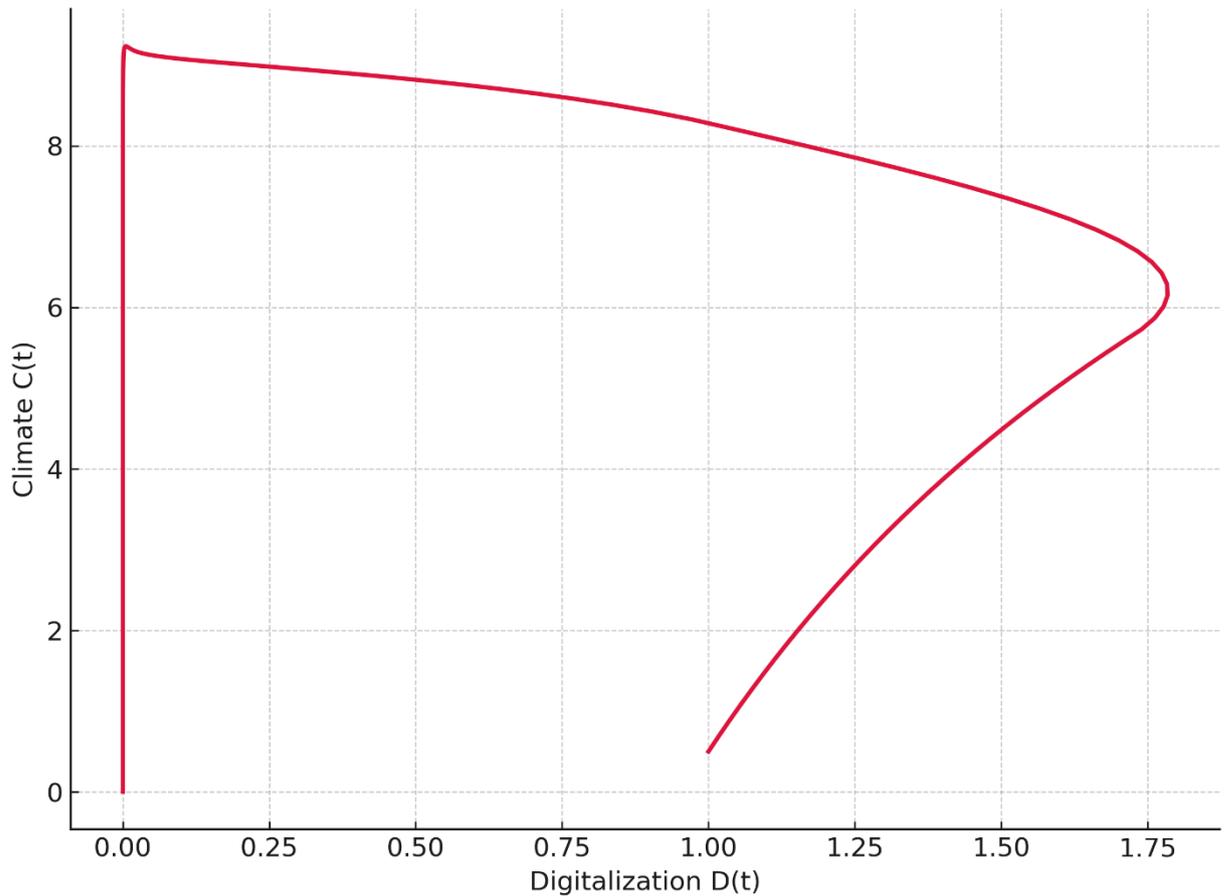

*Figure 2.* Phase portraits: *D(t)* vs. *C(t)*

The phase trajectory starts from the bottom-left and moves upwards and to the right: digitalization *D(t)* increases, whereas climate *C(t)* initially remains low. The trajectory then bends and approaches horizontal saturation—when *C(t)* becomes high enough, the growth of *D(t)* halts. The final shape resembles a homoclinic loop or a slowing growth curve with feedback inhibition.

The graph illustrates that the growth of digitalization itself provokes climatic constraints, which in turn limit further digitalization. This confirms the core hypothesis of the ICF model: a closed dynamic feedback loop exists between the data and climate.

The trajectory neither diverges to infinity nor collapses: it approaches an equilibrium point, indicating the possibility of dynamic system stability.

Under different parameter settings (especially for delays *τ*), the system may shift into cyclical phase loops—analogous to oscillatory patterns in ecological or economic systems.

The phase curve confirms that there is a threshold beyond which climate stress *C(t)* significantly reduces the digitalization growth rate. This is a nonlinear constraint that is invisible in linear models.

The system demonstrates signs of self-organization: the growth of one component induces resistance from another, forming a regulated ecosystem between digital and climatic domains.

The phase curve indicates the presence of a stable equilibrium point between digitalization and climate—achievable under reasonable values of greenness $G(t)$ and controlled growth of $D(t)$.

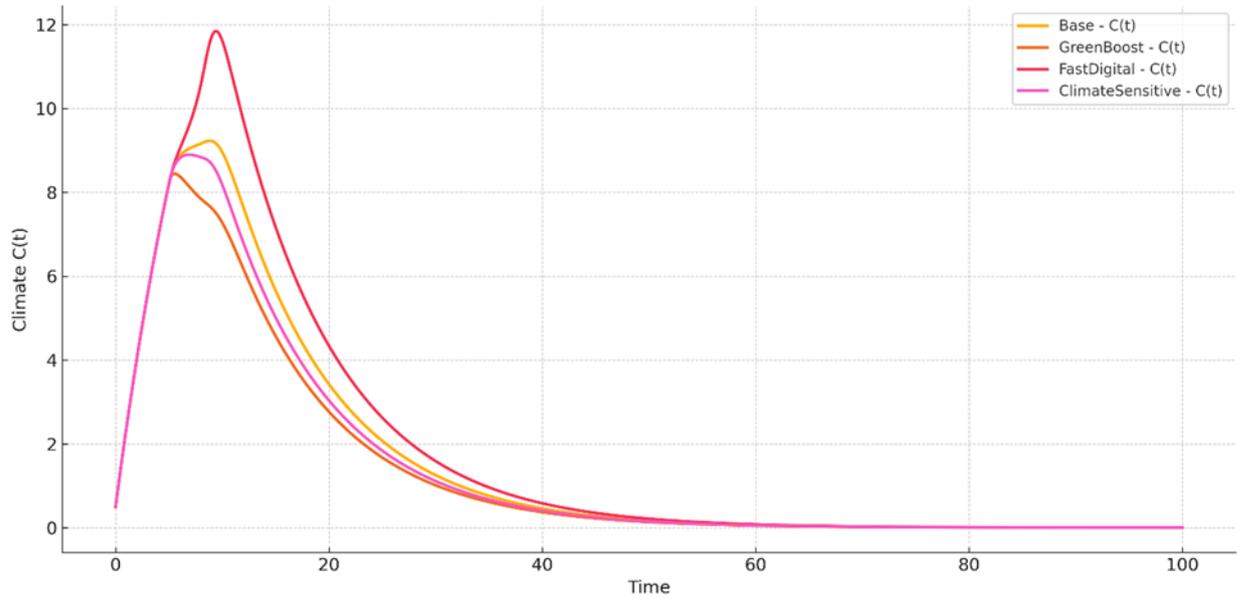

*Figure 3.* Multi-Scenario Trajectories of Climate $C(t)$

*Table 3.* Analytical comparison of scenarios

| Scenario | Final D(t) | Final C(t) | Max C(t) | Final G(t) | G(t) Growth |
|---|---|---|---|---|---|
| Base | 0.00 | 0.00 | 9.23 | 1.00 | 0.80 |
| GreenBoost | 0.01 | 0.00 | 8.45 (↓) | 1.00 | 0.80 |
| FastDigital | 0.00 | 0.00 | 11.85 (↑) | 1.00 | 0.80 |
| ClimateSensitive | 0.00 | 0.00 | 8.90 | 1.00 | 0.80 |

Interpretation of Figure 3 and Table 3:

1. *GreenBoost (accelerated greening):*
    - Lowest climate peak: $C_{max}$=8.45,
    - Rapid growth of greenness effectively mitigates the climate load,
    - The digitalization remains almost intact ($D(t)$=0.01), within numerical accuracy).

2. *FastDigital (aggressive D(t) growth):*

- Highest climate peak: $C_{max}$=11.85,
- Rapid digitalization overloads the system → feedback activates → $D(t)$ collapses to zero,
- Despite rising $G(t)$, greenness cannot compensate for the pace.

3. *ClimateSensitive (vulnerability scenario):*
   - Even moderate digitalization causes strong climate feedback,
   - The system quickly enters the inhibition zone and zeroes $D(t)$,
   - Note that $\epsilon$ (sensitivity) is a critical parameter.

4. *Base (default parameters):*
   - Mid-level climate peak: $C_{max}$=9.23,
   - Predictable and stable behavior under baseline assumptions.

Conclusion from Figures and Table:
- Greenness is key to stabilization—but only if its growth outpaces digitalization.
- Fast digital expansion without adaptation results in collapse by overheating and climate-driven regression.
- Climatic sensitivity $\epsilon$ critically determines system stability: resilient infrastructure must be developed to withstand heat and failure.
- The optimal strategy is controlled digitalization growth + investment in greening.

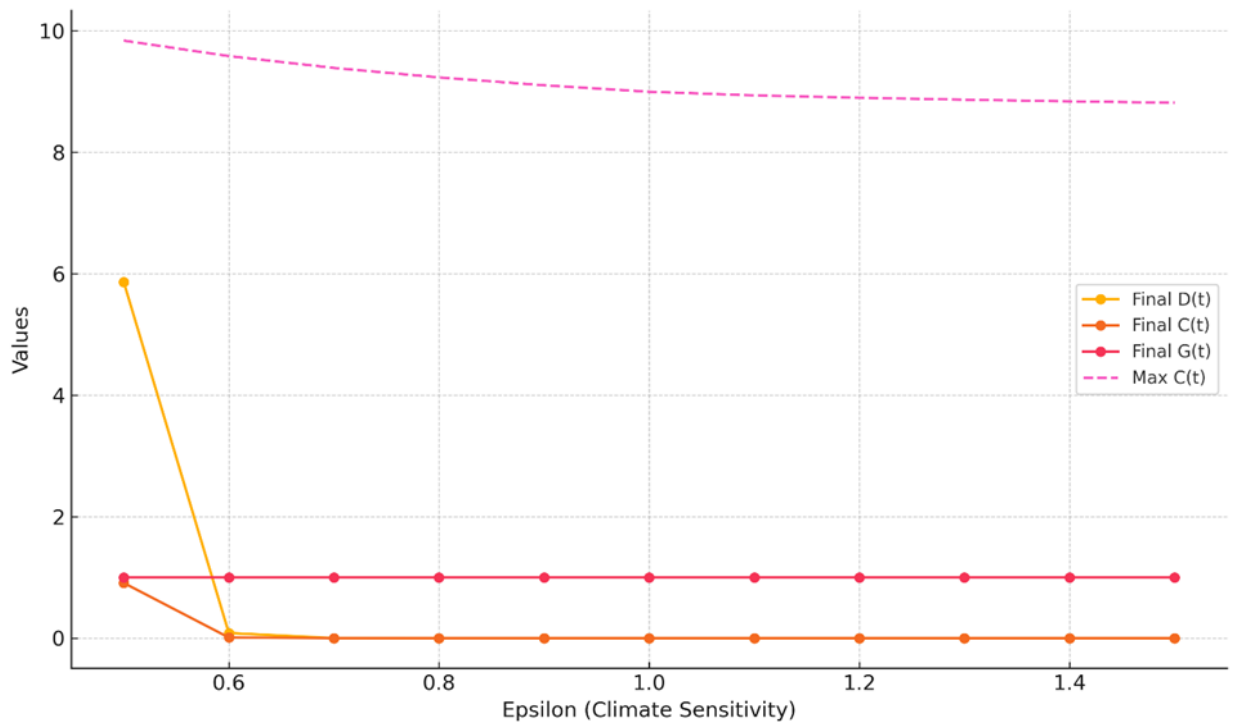

*Figure 4.* Sensitivity Analysis for Parameter $\epsilon$

The parameter $\epsilon$ is interpreted as a structural regulator of digital resilience.

As $\epsilon$ increases, even moderate climate rise leads to the immediate suppression of digitalization growth. The climate peak Cmax increases under high sensitivity, despite the reduction in digital activity.

Parameter $\epsilon$ defines the strength of negative feedback: the higher the temperature *C(t)*, the more digitalization *D(t)* is suppressed.
- At low $\epsilon$, digitalization can continue even under warming conditions.
- At high $\epsilon$, even minor warming severely inhibits or nullifies *D(t)* growth.

Despite the reduction in digitalization, the climate continues to warm—due to the inertia of the thermal response and delay *τ*. Greenness *G(t)* increases to a maximum in all scenarios, but:
- at low $\epsilon$ it successfully mitigates heat,
- at high $\epsilon$, it cannot act in time—because *D(t)* has already been halted.

Thus, parameter $\epsilon$ determines how much the infrastructure "fears" the climate. Even with strong greening, high sensitivity can nullify digitalization.   Excessively high $\epsilon$ leads to a fragile digital ecosystem.

In this mode, digitalization becomes unstable even under short-term climatic fluctuations, and a threshold-like catastrophic collapse may occur.

The optimal band for $\epsilon$ is 0.6–0.9:
- To allow stable digital growth,
- Provides time for green infrastructure to compensate for climate pressure;
- Enables synergetic equilibrium between digitalization and climate.

### 4.7. Political and Regulatory Aspects

Although digitalization appears immaterial, it in fact produces a thermodynamic footprint. This may lead to a delayed catastrophe, as the climate responds slowly but sharply because feedback lags. Accordingly, systemic stability requires regulation of the parameters within the information–climate feedback (ICF) model. Otherwise, a phase breakdown may occur (digital overheating → regulatory emergencies → collapse).

We therefore propose a set of specific policy, regulatory, and strategic measures grounded in the ICF model.

Most importantly, this concerns the creation of a regulatory framework capable of the following:
- containing the climate burden of digitalization,
- incentivizing sustainable growth of the digital infrastructure, and
- building an architecture for preventive digital-climate security.

A concise interpretation of the regulatory structure is given below:

*Table 4*. Key Parameters for Regulation (Based on the ICF Model)

| Parameter | Interpretation | Proposed Regulation |
|---|---|---|
| $\epsilon$ | Sensitivity of IT systems to climate | Introduce a standard of digital thermal stability |
| $\rho$ | Greening rate | Subsidize innovation and green data center adoption |
| $\eta$ | Digitalization growth rate | Establish regional thresholds and quotas for growth |
| $G(t)$ | Real energy efficiency | Implement a Digital Greenness Index (DGI) |

We further propose several normative frameworks within the context of the ICF model:

1. *Digital Thermal Footprint Act*
    - The digital infrastructure (data centers, AI, blockchain) is included in climate quotas.
    - Mandate reporting on energy use, thermal emissions, and compensation via green technologies.
2. *Digital Climate Resilience Index (DCR)*
    - Publish an annual resilience index by country/region as follows:
        - $\epsilon$ (vulnerability),
        - *G(t)* (greenness),
        - *η* (growth rate),
        - *C(t)* (climate response).
3. *Digital Overheating Threshold Zones*
    - A heatmap of hotspots with a high risk of digital-climatic overload was created.
    - In such zones:
        - restrict the construction of new data centers,
        - and mandate passive cooling,
        - introduce a digital density tax.

In the context of a new global strategy — the Green Digital Accord (GDA) — countries must coordinate efforts to limit the climate footprint of digitalization and implement mandatory sustainability standards, including the following:
- Certification of major IT hubs for thermal stability,
- Establishing a global cap on the digital thermal footprint (analogous to the $CO_2$ budget),
- A digital adaptation fund should be created to support high-risk, low-greenness regions.

We outline five policy priorities for ICF governance:

*Table 5.* Five Priorities of the ICF Policy

| Priority | Action Items |
|---|---|
| 1. Contain $\epsilon$ | Thermal resilience standards for infrastructure |
| 2. Accelerate $\rho$ | Subsidies for greening the IT sector |
| 3. Control $\eta$ | Quotas and governance of digitalization growth |
| 4. Monitor $G(t), C(t)$ | National indices and mapping systems |
| 5. Global Coordination | Green Digital Accord as a global digital climate pact |

In terms of a Global Implementation Roadmap for the ICF, we propose the launch and institutionalization of a global agreement aimed at managing the thermal footprint of digitalization, mitigating climate risks, and incentivizing green IT practices.

Roadmap for the Green Digital Accord (GDA)

*Phase I. PREPARATORY STAGE (0–2 years)*

| Step | Action | Responsible Actors |
|---|---|---|
| 1.1 | Formation of the international GDA Initiative Group (via UNFCCC or ITU) | UNEP, ITU, IPCC, UNDP |
| 1.2 | Global audit of the digital thermal footprint | IEA, Greenpeace, IEEE |
| 1.3 | Development of methodologies for DCR, DGI, and DHI (Digital Heat Index) | Scientific Expert Consortium |
| 1.4 | Creation of ICF-based scenario banks | Modelling centers: IIASA, MIT, Potsdam Institute |
| 1.5 | Inclusion of digitalization in the COP climate agenda | UNFCCC, G20 |

*Phase II. AGREEMENT NEGOTIATION (2–4 years)*

| Step | Action | Responsible Actors |
|---|---|---|
| 2.1 | Drafting and publication of the Green Digital Accord | GDA Secretariat |
| 2.2 | Organization of Digital-Climate Summit | UNEP + ITU |
| 2.3 | Negotiation and ratification of limits:<br>– Maximum digital thermal footprint<br>– Digitalization growth rates<br>– Mandatory targets for G(t) | Intergovernmental Groups |

*Phase III. IMPLEMENTATION AND MONITORING (4–10 years)*

| Step | Action | Responsible Actors |
|---|---|---|

| 3.1 | Creation of the Digital Climate Registry — a global platform for monitoring data flows, energy use, and thermal footprint | World Digital Climate Observatory (WDCO) |
| --- | --- | --- |
| 3.2 | National programs for green digitalization:<br>– Quotas for digital growth<br>– Taxes on thermal density<br>– Compensation funds | National Governments |

*Phase IV. INSTITUTIONALIZATION AND UPDATING (10+ years)*

| Step | Action | Responsible Actors |
| --- | --- | --- |
| 4.1 | Update of the agreement based on ICF 2.0 model | GDA Council + Scientific Community |
| 4.2 | Integration of digital-climate metrics into ESG and global investment ratings | Bloomberg, S&P, IFC |
| 4.3 | Introduction of digital green passports for data centers and IT infrastructure | ISO + National Regulators |
| 4.4 | Expansion of the Accord to a full Digital-Environmental Protocol — the 5th pillar of global sustainable development | UNGA, UNDP, G20, WEF |

Final recommendations:

1. Institutionalize the ICF model as the foundation for regulating digital–climate interactions.
2. Develop digital climate passports for regions and infrastructure systems.
3. Standardize key indicators $\epsilon$, $\rho$, $G(t)$, and $C(t)$ as sustainability metrics.
4. Integrate the GDA into climate negotiations and the ESG policy framework.

**5. Discussion**

The ICF model was numerically integrated over a 50-year interval under various initial conditions and levels of climate sensitivity. The results show that digitalization $D(t)$ initially grows exponentially but then slows down due to the climate response $C(t)$, particularly under high values of $\epsilon$. The thermal footprint $H(t)$ accumulates faster than it dissipates, creating inertial overheating. The climate response $C(t)$ exhibited delayed growth, potentially culminating in peak overheating. Greenness $G(t)$ compensates for part of the thermal effect but requires sustained support and investment — otherwise, $G(t) \rightarrow 0$.

On the basis of changes in key parameters, three stable scenarios can be identified:

*Table 6.* Three scenarios of information–climate feedback

| Scenario | Description | Conditions |
| --- | --- | --- |
| A. Sustainable Growth | Digitalization stabilizes; heat is compensated | High G(t), low $\epsilon$, fast greening |
| B. Cyclical Overheating | Alternating phases of growth and inhibition | Medium $\epsilon$, delayed response $\tau$, weak greening |
| C. Digital Collapse | Digitalization drops to minimum; climate destabilizes | High $\epsilon$, delayed response, no compensation |

The nonlinearity of the feedback loop makes the system highly sensitive even to small parameter shifts. Delay $\tau$ amplifies risk: even steadily growing digitalization may "overheat" if the climate reacts too slowly. The compensatory role of *G(t)* can postpone a phase shift but cannot prevent it without sufficient growth. A vulnerability threshold is observed: for $\epsilon > 0.05$, digitalization degrades regardless of the initial conditions.

Our modelling confirms the central hypothesis: information possesses not only energetic value but also thermodynamic risk. On large scales, digitalization behaves as a climate-active agent.

Modern climate and technology policies ignore the effect of delayed overheating, assuming that digitalization is inherently neutral. However, the ICF model shows that even with high energy efficiency, a thermal tipping point may emerge, after which the system degrades.

Note that the aggressive development of green digitalization can reduce *β* (the thermal coefficient) and extend the stability zone but does not eliminate the need to control the growth of *D(t)*.

One of the key outcomes of this study is the reconceptualization of information in the ecological–climatic context. Traditionally, information is treated as an immaterial, abstract resource, but the ICF theory demonstrates that information materializes through infrastructure, produces energy and heat, and thus becomes a new climate-driving factor.

This challenges the existing dichotomy between "clean" digital economies and "dirty" industrial economies, revealing that digitalization does not eliminate exothermic pressure—it masks it with abstractions. In this sense, information becomes a thermodynamically significant entity.

The ICF model reveals a nonlinear, delayed, and self-reinforcing feedback loop:

*Digitalization→Heat→Climate→Pressure on Digitalization*

Unlike classical feedback (albedo, evaporation), this loop is anthropogenic and technological, creating a new contour for risk governance. This loop has not been previously included in climate or technological models. Its structure resembles biological amplification circuits or economic delay loops, thus extending cybernetic interpretations of resilience.

We propose that the ICF loop constitutes a structurally novel form of techno-natural interdependence.

Indeed, current digital policy assumes the following:

"*If digitalization reduces the carbon footprint, it is therefore always sustainable.*"

However, our model reveals this to be a new form of the Jevons paradox: energy efficiency stimulates digital growth, which results in a delayed thermal footprint. The result is a false sense of sustainability, latent risk, and cumulative entropy that cannot be easily neutralized retroactively.

In this context, the ICF framework enables the early detection of hidden overheating thresholds and the simulation of critical trajectories.

Green architecture is a necessary response to digital energy consumption. However, the model shows that $G(t)$ often lags behind $D(t)$, and when $G(t)$ grows too slowly, the compensatory effect is insufficient to counteract heat buildup.

Moreover, green technologies themselves consume resources (rare-earth materials, water, and $CO_2$ emissions). Thus, greenness is necessary but not sufficient. We must also regulate the growth rate of digitalization itself. Tactically, this requires synergy between green innovation and temporal governance.

The occurrence of phase transitions is especially critical. Under certain values of $\epsilon$ and $\tau$, the system loses stability; digitalization decreases toward collapse, and even high $G(t)$ values cannot compensate without strategic control over $D(t)$. This brings the ICF closer to catastrophe theory than linear forecasting does. The assumption of smooth digital growth is a dangerous illusion.

Therefore, a strategy of climatic inertial buffering for digital transitions is needed.

The thermal footprint of information is spatially uneven. Developed countries (the US, China, and the EU) constitute 80% of data centers (Energy and AI, 2025), AI models, and digital chains. Our ICF model indicates that advanced digital clusters export thermal risks, whereas the Global South merely absorbs the climatic consequences without being a subject of digital architecture.

This raises the issue of climate justice, not only in carbon terms but also in informational terms. Hence, a global agreement on the fair distribution of the digital climate burden is needed—the Green Digital Accord.

We propose a table of key metrics and indicators for international coordination under the Green Digital Accord—a normative framework for sustainable digitalization informed by the ICF model.

*Table 7*. Sustainable Digitalization Metrics for the Green Digital Accord

| No. | Metric/Indicator | Units | Relevance for GDA |
|---|---|---|---|
| 1 | Digital Thermal Density | $W/m^2$ or $J/m^3$ | Thermal load per unit of infrastructure area |
| 2 | Infoclimatic Sensitivity | $\Delta°C$ per $\Delta D$ | Degree to which digitalization affects climate |
| 3 | Digitalization Growth Rate | % per year | Key risk factor for overheating |
| 4 | Greening Rate | % per year | Must ≥ growth rate of $D(t)D(t)D(t)$ |
| 5 | Total Digital Thermal Footprint | TJ, PJ | Cumulative heat from IT infrastructure |
| 6 | Thermal Efficiency of IT Systems | % | Useful energy relative to waste heat |
| 7 | Digital Greenness Coefficient | 0 to 1 | Share of digital systems compensated by renewables |
| 8 | Climate Inertia Time | years | Delay between digital activity and climate response |
| 9 | Climate Vulnerability Index | 0–1 or score | Susceptibility of data centers to climate |
| 10 | Global Distributed Thermal Map | heatmap | Inequality of thermal impact across countries |
| 11 | Digital Sustainability Entropy | bits/°C or index | Digital volume relative to climate stability |
| 12 | Digital Climate Justice Index | index (0–1) | Equity of thermal burden among nations |
| 13 | Informational Energy Intensity | J/bit | Energy required to process a single bit |
| 14 | Recovered Heat Ratio | % | Proportion of heat that is reused or recovered |
| 15 | Digital Overheating Threshold | index (bits/capita/year) | Maximum safe level of digital activity |
| 16 | ICF Reporting Compliance | binary (0/1) or score | Presence of national ICF reporting standards |

These indicators can be used for country-level and corporate monitoring (e.g., Google, Amazon, Microsoft, and Baidu), for setting thermal density quotas and growth caps, and forming a "stability zone" in global climate policy — not only for $CO_2$ but also for the thermal contribution of digital systems.

In summary, we reiterate the following conclusions:
- This information creates thermodynamic feedback.
- The ICF model demonstrates that digitalization, despite its intangible appearance, has a material climatic footprint via accumulated heat and infrastructure energy demand.
- Digitalization can become climatically unstable. Phase breakdowns occur under high $\epsilon$, slow $G(t)$, or large delays $\tau$.
- Green digitalization is necessary but not sufficient. Compensation works only if $G(t)$ grows faster than $D(t)$.

- A new category of climate assessment is needed — the infoclimatic footprint. Current sustainability protocols (IPCC, ISO) ignore the impact of IT loads on the climate.
- The ICF should be adopted as a new paradigm for sustainable digital planning—bridging thermodynamics, cyber-physical systems, and ecology into a unified model.
  Recommendations:
- Integrate the thermal footprint of digitalization into global climate models (e.g., CMIP7).
- Empirical studies on temperature–load–failure correlations in data centers have been conducted.
- Development of a theoretical foundation for infothermodynamics as an interdisciplinary field.
- The Green Digital Accord, analogous to the Kyoto Protocol, was drafted for the digital climate burden.
- The introduction of national quotas on the infoclimatic load, especially for megaplatforms, is needed.
- Establish ISO/IEEE/ITU standards for green digital architecture.
- The thermal mapping and reporting for data centers should be mandated.
- Implement adaptive thermal managers in AI infrastructure that is responsive to the external climate.
- Invest in heat-neutral technologies: passive cooling, bioenergy compensators.
- The ICF paradigm should be included in the ecological curricula and digital professions.
- Development of open-source ICF simulators for public monitoring and education.
- Raise awareness of informational climate injustice — where digital activity in the Global North impacts the climate of the Global South.
- Ethical indices for digital infrastructure, including thermal impact, local harm, and resource intensity, should be established.

**6. Conclusion**

In the era of rapidly expanding digital civilization, humanity faced, for the first time, the need to recognize immaterial information as a material climatic agent. The information–climate feedback (ICF) theory, substantiated in this article, demonstrates that digitalization—contrary to its widespread perception as a "clean" force—generates a real thermal and energetic footprint capable of influencing the global climate system.

This impact is nonlinear, nonlocal, and noninstantaneous. It accumulates, unfolds inertially over time, and forms hidden feedback loops in which:

*Information→Infrastructure→Heat→Climate→Digitalization*

—forming a closed systemic loop.

The nonlinear dynamic ICF model presented in this study not only formalizes this interdependence but also identifies phase thresholds beyond which the system loses stability. These thresholds are tied to the growing mismatch between the pace of digital growth and the capacity of green compensatory infrastructure to neutralize its effects.

In other words, not all digitalization is sustainable: when the balance between informational and thermodynamic circuits is disrupted, we risk triggering a digitally induced climate shift.

In addition to its scientific significance, ICF theory also has broad political and regulatory implications. It points to the need for the following:
- Including the infoclimatic footprint in climate reports and impact assessments,
- By introducing global quotas for digital thermal loads,
- Establishing an international Green Digital Accord,
- Development of ethical and climate indices for major IT corporations.

The issue of digital climate justice is especially pressing: developed countries concentrate on capacity, dominate data centers, and scale up AI systems, whereas developing countries remain more vulnerable to the climate effects they do not generate. This constitutes a new form of global digital–ecological inequality.

From a philosophical standpoint, the ICF theory breaks down the conventional dichotomy between "information" and "matter," showing that every form of information exists on a thermodynamically burdened substrate. Every act of transmission, storage, or computation has an energetic and thermal cost.

This calls not only for new technical solutions but also for a new understanding of the limits of digital growth, the governance of informational space, and the ethics of technological expansionism.

In this context, green digitalization is not only a trend or an ESG label but also a historical necessity for adapting digital civilization to the planetary limits of sustainability. Only through the integration of information, climate, energy, and ethics can we ensure not a collapse but rather a coevolution of digital and biospheric futures.

Thus, digitalization is not merely a question of progress; it has become a question of survival. We argue that every byte must now be reconciled with the biosphere.

The development of global digital infrastructure, which has long been assumed to be inherently progressive and "clean," has taken on a new thermodynamic and climatic dimension in light of ICF theory. Digitalization can no longer be viewed outside the context of ecosystem sustainability: each byte of information carries a potential thermal footprint; each data center contributes to local climate stress; and each unit of growth increases the risk of systemic disruption.

The nonlinear dynamic ICF model reveals a structural feedback loop between data, energy, heat, and climate, identifying critical scenarios in which unchecked digitalization undermines itself through climate destabilization.

This compels a revision of sustainable development policies, where digital transformation can no longer remain outside the scope of climate governance.

By introducing the concepts of infoclimatic sensitivity, information thermal density, green compensation, and digital phase transition, we establish a foundation for a new scientific and political paradigm — where green digitalization is no longer a slogan but an existential necessity.

Accordingly, ICF theory not only diagnoses the invisible risks of digital infrastructure but also offers predictive, evaluative, and regulatory tools, including international agreements, norms, and standards.

In conclusion, green digitalization is not simply a technical adjustment but rather a new survival strategy for the digital–climatic epoch.